\documentclass[conference,10pt]{IEEEtran}
\usepackage{epsfig,epsf,rotating,setspace,latexsym,amsmath,amssymb,amsfonts,bm,theorem,subfigure,epstopdf}
\usepackage{cite,authblk}
\usepackage{bbm}
\usepackage{algorithm}
\usepackage[noend]{algpseudocode}
\usepackage{color}
\usepackage{mathtools}
\usepackage{soul}

\algrenewcommand\algorithmicforall{\textbf{foreach}}
\algrenewcommand\algorithmicindent{.8em}

\IEEEoverridecommandlockouts
\allowdisplaybreaks

\begin{document}
 
\title{Susceptibility of Age of Gossip to Timestomping}
 
\author{Priyanka Kaswan \qquad Sennur Ulukus\\
        \normalsize Department of Electrical and Computer Engineering\\
        \normalsize University of Maryland, College Park, MD 20742\\
        \normalsize  \emph{pkaswan@umd.edu} \qquad \emph{ulukus@umd.edu}}
 
\maketitle

\begin{abstract}
We consider a fully connected network consisting of a source that maintains the current version of a file, $n$ nodes that use asynchronous gossip mechanisms to disseminate fresh information in the network, and an adversary who infects the packets at a target node through data \emph{timestamp manipulation,} with the intent to replace circulation of fresh packets with outdated packets in the network. We show that a single infected node increases the expected age of a fully connected network from $O(\log n)$ to $O(n)$. Further, we show that the optimal behavior for an adversary is to reset the timestamps of all outgoing packets to the current time and of all incoming packets to an outdated time. Additionally, if the adversary allows the infected node to accept a small fraction of incoming packets from the network, then a large network can manage to curb the spread of stale files coming from the infected node and pull the network age back to $O(\log n)$. Lastly, we show that if an infected node contacts only a single node instead of all nodes of the network, the system age can still be degraded to $O(n)$. These show that fully connected nature of a network can be both a benefit and a detriment for information freshness; full connectivity, while enabling fast dissemination of information, also enables fast dissipation of adversarial inputs.
\end{abstract}

\section{Introduction}

Sensor networks generally have limited resources, which prevents them from implementing traditional computer security techniques, making them vulnerable to adversarial attacks. Uncertain dynamics of such networks often force them to rely on decentralized \emph{gossip protocols} \cite{Demers1987EpidemicAF,Minsky02cornellthesis, vocking2000, Pittel1987OnSA, deb2006AlgebraicGossip, devavrat2006, Sanghavi2007GossipFileSplit, amazondynamo, Cassandra, Yates21gossip_traditional,Yates21gossip, baturalp21comm_struc, Bastopcu21gossip, kaswan22slicingcoding} for information dissemination, where information is exchanged between nodes repeatedly and asynchronously using their local status. Gossip protocols were  introduced and have been widely used in the context of distributed databases. In this work, we consider the presence of an adversary in a gossip network \cite{Nguyen17interferencegame, Garnaev19jamming, Xiao18jamming, kaswan22jammerring, Banerjee22adversary, Banerjee22game, Augustine16_adversaries, Georgiou08_complexitygossip}, who corrupts the gossip operation by manipulating the timestamps of some data packets flowing in the network, a technique known as \emph{timestomping} \cite{minnaard2014timestomping}, with the goal of bringing about staleness and inefficiency to the network. A timestomping attack can be launched in many ways. For instance, a malicious insider node can deviate from the gossip protocol and inject old packets by rebranding them as fresh packets via timestamp manipulation, while maintaining the gossiping frequency to evade suspicion. Other methods include \emph{meddler in the middle} (MITM) attacks, where the adversary inserts its node undetected between two nodes and manipulates communication, and \emph{eclipse} attacks  where the adversary manipulates the target node by redirecting its inbound and outbound links away from legitimate neighboring nodes to adversary controlled nodes, thereby isolating the node from the rest of the network, as encountered in gossip based blockchain networks.

Most prior works on gossip networks consider total dissemination time of a message in the network as the performance metric. For instance, \cite{vocking2000} shows that dissemination of a single rumor to $n$ nodes takes $O(\log n)$ rounds, \cite{deb2006AlgebraicGossip} shows that $n$ messages can be disseminated to $n$ nodes in $O(n)$ time in fully connected networks using random linear coding (RLC), \cite{devavrat2006} provides an analogous result for arbitrarily connected graphs, and  \cite{Sanghavi2007GossipFileSplit} analyzes dissemination of messages by dividing them into pieces. However, highly dynamic nature of data sources in modern applications prevents these networks from waiting for a specific message to reach all nodes of the network before fresh information can be circulated. Distributed databases \cite{amazondynamo, Cassandra}, for example, employ timestamp versioning, wherein every new information is created with a timestamp value taken from the system clock. When two nodes come in contact to exchange information, the timestamps of data at both nodes are compared and the node carrying the data with older timestamp discards its data for the fresher data of the other node. 

In this regard, age of information \cite{Kosta17agesurvey, Sun19agesurvey, yates21agesurvey} may be a more suitable indicator of network efficiency. Given $U_i(t)$ as the timestamp of the packet with node $i$ at time $t$, the instantaneous age of information is given by $X_i(t)=t-U_i(t)$. The nodes wish to have access to the most up-to-date information at all times, and therefore, are prompted to decrease $X_i(t)$ by fetching packets with more recent timestamps, e.g., with higher $U_i(t)$. Gossip networks have been studied from timeliness perspective in \cite{Yates21gossip_traditional, Yates21gossip, baturalp21comm_struc, Bastopcu21gossip, kaswan22slicingcoding}. \cite{Yates21gossip_traditional, Yates21gossip} derive the recursive age equations using stochastic hybrid system framework for age, \cite{baturalp21comm_struc} studies the expected version age in clustered gossip networks,  \cite{Bastopcu21gossip} extends these results to the binary freshness metric, \cite{kaswan22slicingcoding} considers age scaling in gossip networks using file slicing and network coding, and \cite{kaswan22jammerring} studies the effects of jamming adversaries on gossip age in ring networks. 

Timestomping is often used by malware authors as an anti-forensics technique to make files blend in with the rest of the system. In this work, an adversary uses timestomping with the goal of worsening the expected age in the network. Consider two nodes, $A$ and $B$, that randomly come in contact to exchange information and consider the presence of adversary at node $A$ capable of altering timestamps of all incoming and outgoing files. If node $A$ is outdated compared to node $B$, the adversary would be inclined to increase the timestamp of an outgoing packet from node $A$ to make it appear fresher so as to misguide node $B$ into discarding its packet in favor of a staler packet, and also, decrease the timestamp of an incoming packet from node $B$ so as to avoid its acceptance at node $A$. Conversely, if node $A$ is more up-to-date than node $B$, the adversary would reduce timestamps of outgoing packets and increase timestamps of incoming packets to make node $B$ reject fresher files and node $A$ accept staler files. More the manipulated timestamps digress away from their true value, higher are the chances of error in deciding which packet should be discarded, since this decision is based on a comparison of timestamps. At time $t$, the maximum error is caused when file timestamp is either changed to the current time $t$ or the earliest time $0$. Thus, we consider an adversary, who, for each packet, makes the decision of changing its timestamp to either $t$ or $0$. The adversary is \emph{oblivious} in that it does not look into a packet and see its actual timestamp. Thus, the adversary changes the timestamp to either $t$ or $0$ probabilistically.

In this paper, we consider a gossip network where an adversary captures a node and manipulates the timestamps of packets coming into and going out of the node (see Fig.~\ref{fig:fully_conn_net_eclipse_attack}). We show that one infected node can single-handedly suppress the availability of fresh information in a large network of $n$ users employing a gossip protocol, and increase the expected age in a complete graph from $O(\log n)$ found in \cite{Yates21gossip_traditional} to $O(n)$. In addition, we show that the optimal action for the adversary is to always increase the timestamp of every outgoing packet to $t$ and decrease the timestamp of every incoming packet to $0$, in effect, preventing all incoming files from being accepted and actively persuading other nodes to always accept outgoing packets from the infected node. Further, we show that if the the infected node is allowed to accept even a small fraction of incoming packets from the network, then a large network can curb the spread of stale files coming from the infected node by effectively lowering the infected node's age. These observations show how the fully connected nature of a network can be both a benefit and a detriment for network staleness. Additionally, we show that if the malicious node contacts only one other node instead of all nodes of the network (see Fig.~\ref{fig:fully_conn_net_MITM_attack}), the system age can still be degraded to $O(n)$, which highlights how little an effort is needed on the part of the adversary to bring down the freshness of the entire network. 

\section{System Model and SHS Characterization}

\begin{figure}[t]
\centerline{\includegraphics[width=0.85\linewidth]{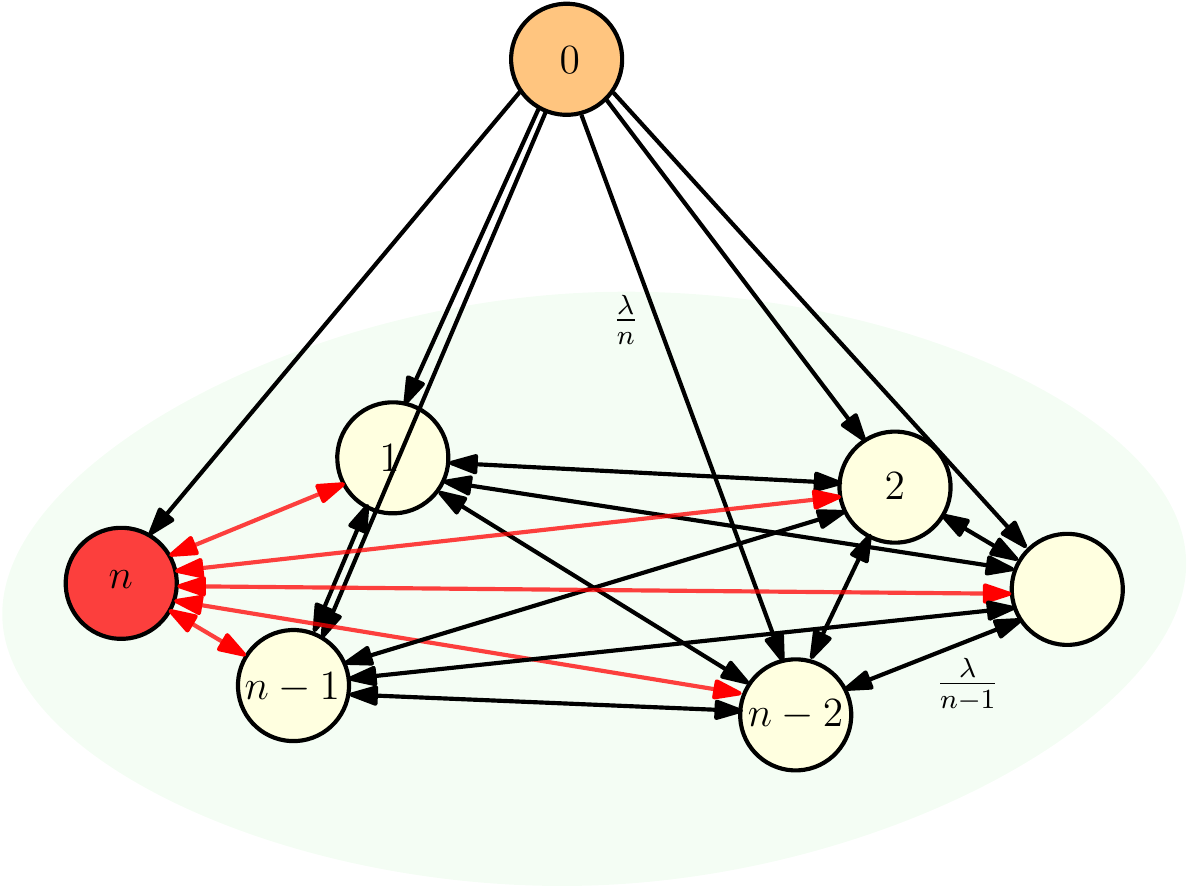}}
\caption{Fully connected network of $n$ nodes with an infected node.}
\label{fig:fully_conn_net_eclipse_attack}
\vspace*{-0.4cm}
\end{figure}

We study a fully connected network, shown in Fig.~\ref{fig:fully_conn_net_eclipse_attack}, which comprises a source and $n$ user nodes $\mathcal{N}=\{1,\ldots,n\}$. The source, alternatively referred to as node $0$, is assumed to always posses the latest file packet and consequently has zero age at all times. The nodes wish to acquire the most up-to-date file to lower their average age from the source, who updates each user node as a Poisson process with rate $\frac{\lambda}{n}$. Further, a user node $i$ randomly sends its current packet to a user node $j$ according to a Poisson process with rate $\lambda_{ij}=\frac{\lambda}{n-1}$. Thus, all nodes send out updates after exponential inter-update times with a total rate $\lambda$. Let $U_j(t)$ denote the timestamp marked on the file stored at the node $j$. Then, at the receiving node $j$, the claimed timestamp of the incoming packet is compared with $U_j(t)$ to determine which packet should be kept. Note that a node always accepts an update from the source which generates update packets with current timestamp $t$. 

We assume that the highest index node, node $n$, is under attack by an adversary that manipulates the timestamps of all incoming and outgoing packets of node $n$. For the outgoing packets, the adversary chooses to increase the timestamp (to current time $t$) with probability $p$ and decrease the timestamp to $0$ with probability $1-p$. Similarly, the adversary increases and decreases the timestamp of incoming packets with probability $1-p$ and $p$, respectively. We will refer to the nodes in set $\mathcal{N}_R=\{1,\ldots,n-1\}$ as \emph{regular} nodes and node $n$ as the infected node. We assume that the infected node always accepts packets from the source like other nodes, delivered to it with rate $\frac{\lambda}{n}$, which helps the adversary evade suspicion of malicious activity by maintaining a remote contemporary relevance of the contents of its manipulated packets.

We denote the long-term average age at node $i$ by $v_i$, where $v_i=\lim_{t \to \infty} \mathbb{E}[X_i(t)]$, and wish to study its extent of deterioration through timestomping. Note that the actual instantaneous age at node $i$ is $X_i(t)= t-\bar{U}_i(t)$, where  $\bar{U}_i(t)$ indicates the true packet generation time, which can be different from the claimed timestamp $U_i(t)$ if the file timestamp has been tampered with. For a set of nodes $S$ at time $t$, let $X_{N(S)}(t)$ indicate the actual instantaneous age of the node claiming to possess the most recent timestamped packet in set $S$, i.e., $X_{N(S)}(t)=X_{\arg \max_{j\in S} U_j(t)}(t)$. We define $v_S=\lim_{t \to \infty} \mathbb{E}[X_{N(S)}(t)]$. Here we would like to point out that in a network without adversary where all files are marked with true timestamps, $X_{N(S)}(t)$ reduces to $X_S(t)= \min_{j\in S}X_j(t)$ defined in \cite{Yates21gossip_traditional}, since the node with highest timestamp will also have the lowest age in the set $S$. 

Reference \cite{Yates21gossip_traditional} demonstrates how stochastic hybrid system (SHS) models yield linear equations useful for deriving long-term average age at nodes in a gossip network of $n$ users with a given topology. Due to the presence of a timestomping adversary, we choose the continuous state for our SHS model as $(\pmb{X}(t),\pmb{U}(t))\in \mathbb{R}^{2n}$, where $\pmb{X}(t)=[X_1(t),\ldots,X_n(t)]$ denotes the instantaneous ages at the $n$ nodes and $\pmb{U}(t)=[U_1(t),\ldots,U_n(t)]$ denotes the timestamps marked on the packets at the $n$ nodes at time $t$. The convenience of the SHS based age characterization follows from the presence of a single discrete mode with trivial stochastic differential equation $(\pmb{\dot X}(t),\pmb{\dot U}(t))=(\pmb{1}_n,\pmb{0}_n)$, where the age at each node grows at unit rate when there is no update transfer, since the timestamps of the node packets do not change between such transitions. Consider a test function $\psi:\mathbb{R}^{2n}\times [0,\infty) \to \mathbb{R}$ that is time-invariant, i.e., its partial derivative with respect to $t$ is $\frac{\partial\psi(\pmb{X},\pmb{U},t)}{\partial t}=0$, such that we are interested in finding its long-term expected value $\mathbb{E}[\psi]=\lim_{t \to \infty} \mathbb{E}[\psi(\pmb{X}(t),\pmb{U}(t),t)]$. Since the test function only depends on the continuous state values $(\pmb{X},\pmb{U})$ and is time-invariant, for simplicity, we will drop the third input $t$ and write $\psi(\pmb{X},\pmb{U},t)$ as $\psi(\pmb{X},\pmb{U})$, which we assume to flow according to the differential equation $\dot \psi(\pmb{X}(t),\pmb{U}(t))=1$. Let $\mathcal{L}$ correspond to the set of directed edges $(i,j)$, such that node $i$ sends updates to node $j$ on this edge according to a Poisson process of rate $\lambda_{ij}$, with this transition resetting the state $(\pmb{X},\pmb{U})$  at time $t$ to $\phi_{i,j}(\pmb{X},\pmb{U},t)\in \mathbb{R}^{2n}$ post transition. Defining $\mathbb{E}[\psi(\phi_{i,j})]=\lim_{t \to \infty} \mathbb{E}[\psi(\phi_{i,j}(\pmb{X}(t),\pmb{U}(t),t))]$, \cite[Thm.~1]{hespanhashs} yields 
\begin{align} \label{eqn:hespanha_eqn}
    0=1+\sum_{(i,j)\in \mathcal{L}}\lambda_{ij}(\mathbb{E}[\psi(\phi_{i,j})]- \mathbb{E}[\psi] )
\end{align}
which is similar to derivations in \cite{Yates21gossip_traditional}, where the left side becomes $0$ as expectations stabilize. We will be using this equation repeatedly by defining a series of time-invariant test functions appropriate for our analysis. For more details, the reader is encouraged to look at references \cite{hespanhashs} and \cite{Yates21gossip_traditional}.

\section{Age Scaling in the Presence of an Adversary} \label{sect:no-capture}
Note that packets arriving at infected node $n$ from a node $i\in \mathcal{N}_R$ with rate $\frac{\lambda}{n-1}$ Poisson process are accepted (or discarded) with probability $1-p$ (or $p$) when the adversary changes timestamp of incoming packet to $t$ (or $0$) to make it appear fresh (or stale). This is equivalent to packets arriving at node $n$ from node $i$ with thinned Poisson process with rate $\lambda_{in}=\frac{(1-p)\lambda}{n-1}$ such that these packets are always accepted. The remaining packets are always discarded and have no effect on age dynamics of the system. Similarly, as the outgoing packets from the infected node $n$ are accepted at node $i\in \mathcal{N}_R$ with probability $p$, this is equivalent to node $n$ sending packets with timestamp $t$ to node $i$ with a thinned Poisson process of rate $\lambda_{ni}=\frac{p\lambda}{n-1}$ such that these packets are always accepted.

Therefore, based on transition $(i,j)$ at time $t$, the reset map $\phi_{i,j}(\pmb{X},\pmb{U},t)=[X_1',\ldots,X_n',U_1',\ldots,U_n']$ can be described by
\begin{align}\label{eqn:eclipse_U_resetmap}
U_{\ell}' = \begin{cases} 
t, & i=0,j\in \mathcal{N},\ell=j\\
\max\{U_i,U_{\ell}\}, & i,j\in \mathcal{N}_R,\ell=j\\
t, & i=n,j\in \mathcal{N}_R,\ell=j\\
t, & i\in \mathcal{N}_R,j=n,\ell=j\\
U_{\ell}, & \text{otherwise}
\end{cases}
\end{align}
and
\begin{align} \label{eqn:X_reset}
X_{\ell}' = \begin{cases} 
0, & i=0,j\in \mathcal{N},\ell=j\\
X_{N(\{i,\ell\})}, & i,j\in \mathcal{N}_R,\ell=j\\
X_n, & i=n,j\in \mathcal{N}_R,\ell=j\\
X_i, & i\in \mathcal{N}_R,j=n,\ell=j\\
X_{\ell}, & \text{otherwise}
\end{cases}
\end{align}
Here, $X_{N(S)}= X_{\arg \max_{j\in S}U_j}$ for state $(\pmb{X},\pmb{U})$ and a subset of nodes $S$. Since all regular nodes have statistically similar age processes, every arbitrary set $S_k$ of $k$ regular nodes will have the same expected age $v_{S_k}$, $S_k \subseteq \mathcal{N}_R$, with $v_{S_1}=v_1$. We pick our first test function to be $\psi(\pmb{X},\pmb{U})=X_{N(S_k)}$, which is modified upon transition $(i,j)$ to  $\psi(\phi_{i,j}(\pmb{X},\pmb{U},t))=X_{N(S_k)}'$. This in turn is characterized using (\ref{eqn:eclipse_U_resetmap}) and (\ref{eqn:X_reset}) as 
\begin{align} \label{eqn:X_NS_k_reset}
X_{N(S_k)}' = \begin{cases} 
0, & i=0,j\in S_k\\
X_{N(S_{k}\cup\{i\})}, & i\in \mathcal{N}_R\backslash S_k,j\in S_k\\
X_n, & i=n,j\in S_k\\
X_{N(S_k)}, & \text{otherwise}
\end{cases}
\end{align}
Noting that $\lambda_{ij}$ is $\frac{\lambda}{n}$ when $i = 0$, and it is $\frac{\lambda}{n-1}$ when $i,j\in \mathcal{N}_R$ and considering the thinned Poisson processes related to node $n$, using (1), this test function yields,
\begin{align}\label{eqn:eclipse_attack_hespanha_equation}
    0=&1+\frac{k\lambda}{n}(0-v_{S_k})+\frac{(n-k-1)k\lambda}{n-1}(v_{S_{k+1}}-v_{S_k}) \nonumber \\
    &+\frac{kp\lambda}{n-1}(v_{n}-v_{S_k}) 
\end{align}
which upon rearrangment gives
\begin{align}\label{eqn:eclipse_vSk}
    v_{S_k}=\frac{\frac{1}{k\lambda}+\frac{n-k-1}{n-1}v_{S_{k+1}}+\frac{pv_n}{n-1}}{\frac{1}{n}+\frac{n-k-1}{n-1}+\frac{p}{n-1}}
\end{align}
Our second test function is simply $\psi(\pmb{X},\pmb{U})=X_n$, i.e., the age at infected node, such that its $(i,j)$ transition map is 
\begin{align} \label{eqn:X_n_reset}
X_n' = \begin{cases} 
0, & i=0,j=n\\
X_i, & i\in \mathcal{N}_R,j=n\\
X_n, & \text{otherwise}
\end{cases}
\end{align}
which, upon proceeding similarly to (\ref{eqn:eclipse_attack_hespanha_equation}) and (\ref{eqn:eclipse_vSk}), gives
\begin{align}\label{eqn:Xn_eclipse_adversary}
    v_n=\frac{\frac{1}{\lambda}+(1-p)v_1}{\frac{1}{n}+(1-p)}
\end{align}

Our goal is to obtain an analytical expression for expected age of a regular node $v_{S_1}=v_1$, by making use of (\ref{eqn:eclipse_vSk}) and (\ref{eqn:Xn_eclipse_adversary}). 

\subsection{Case 1: $p=1$}
In this case, the adversary blocks all incoming packets from the regular nodes, and misleads them into accepting all packets sent by infected node $n$ through timestamp manipulation. Let $y_k=v_{S_k}\frac{n-k}{n-1}$ and using $\frac{1}{n-1}\approx \frac{1}{n}$ for large $n$, (\ref{eqn:eclipse_vSk}) becomes
\begin{align}
    y_k=\frac{n-k}{n-k+1}\bigg( y_{k+1}+ \frac{1}{k\lambda} +  \frac{v_n}{n-1} \bigg)
\end{align}
Starting from $y_1=v_1$, and successively substituting for $y_2, y_3, \ldots, y_{n-1}$, we obtain
\begin{align}
    v_1=& \frac{1}{\lambda} \sum_{k=1}^{n-1} \frac{n-k}{nk} +v_n\sum_{k=1}^{n-1}\frac{n-k}{n(n-1)} \\
    =& \frac{1}{\lambda} \sum_{k=1}^{n-1}\frac{1}{k} -\frac{1}{\lambda}\frac{n-1}{n} +\frac{v_n}{n(n-1)}\sum_{k=1}^{n-1} k \\
    =& O(\log n) + \frac{v_n}{2} \label{eqn:eclipse_iteratively}
\end{align}
since $\sum_{k=1}^{n-1}\frac{1}{k}$ grows asymptotically as $\log n$ and $\frac{n-1}{n} \approx 1$. The $v_n$ in the second term can in turn be obtained by substituting $p=1$ in (\ref{eqn:Xn_eclipse_adversary}), giving $v_n=\frac{n}{\lambda}$.
Hence, 
\begin{align} \label{eqn:v1_vn_by_2_case1}
    v_1= O(\log n) + \frac{n}{2\lambda}=O(n)
\end{align}
To put this deterioration in age scaling into perspective, remember that in a fully disconnected network with no gossiping \cite{baturalp21comm_struc}, expected age at each node also scales as $O(n)$, to be exact, $\frac{n}{\lambda}$, a fact that will come handy later.

\subsection{Case 2: p=0}
In this case, the infected node accepts all files from the $n-1$ regular nodes but does not transmit any files, even when it possesses the latest file, thereby limiting its contribution, positive or negative, to the system age. 

Substituting $p=0$ in (\ref{eqn:eclipse_vSk}) and assuming $\frac{1}{n-1}\approx \frac{1}{n}$ for large $n$ gives $y_k=\frac{1}{k\lambda}+y_{k+1}$ which, upon solving iteratively for $k=\{1,\ldots,n-1\}$, gives $v_1=\frac{1}{\lambda} \sum_{k=1}^{n-1}\frac{1}{k}=O(\log n)$. Hence, for large $n$, such an adversary has negligible effect on the network age. In addition, putting $p=0$ in (\ref{eqn:Xn_eclipse_adversary}) gives 
\begin{align}\label{eqn:eclipse_vn_case2}
    v_n=\frac{\frac{1}{\lambda}+v_1}{\frac{1}{n}+1} \approx \frac{1}{\lambda}+v_1
\end{align}
Hence $v_n$ scales as $O(\log n)$ similar to $v_1$, which indicates that the degradation in age caused by adversary upon increasing timestamps of incoming files is negligible.

\begin{figure}[t]
\centerline{\includegraphics[width=0.85\linewidth]{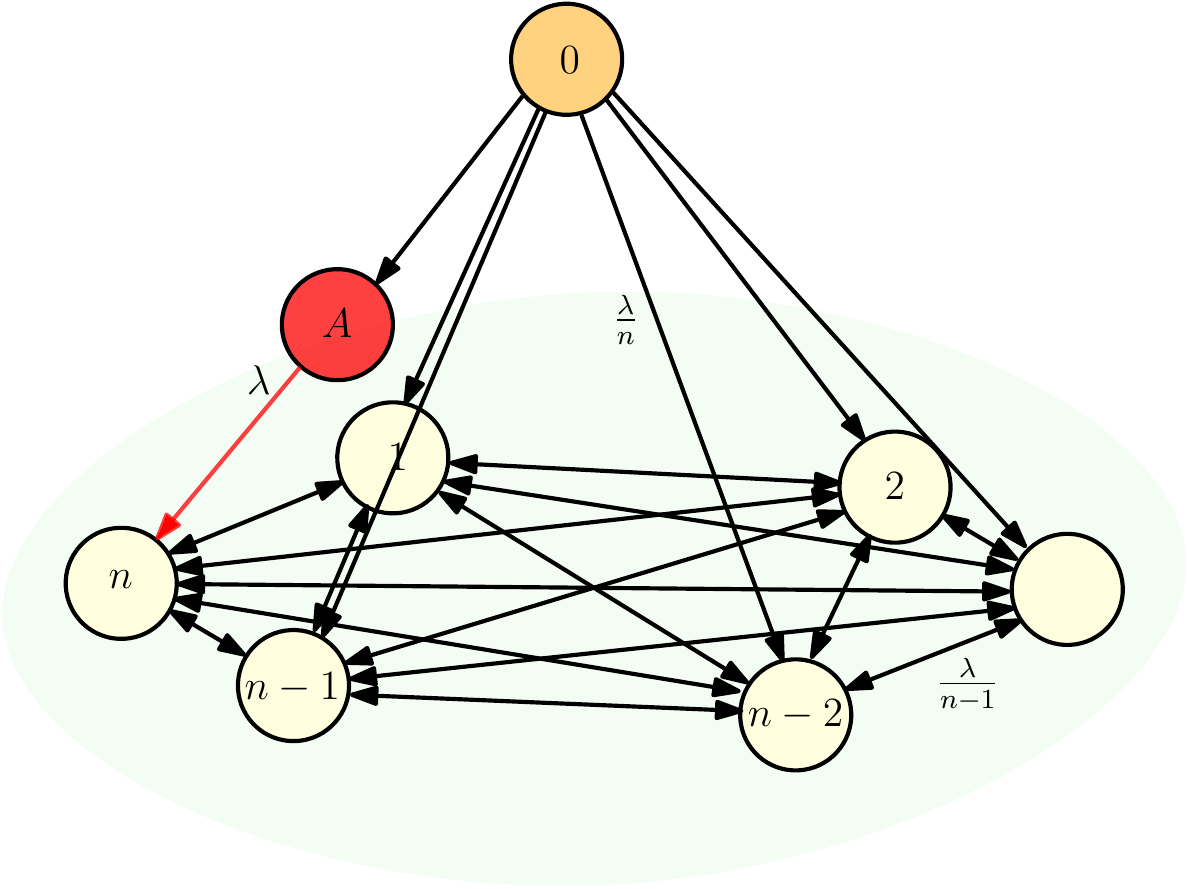}}
\caption{MITM attack on fully connected network of $n$ nodes.}
\label{fig:fully_conn_net_MITM_attack}
\vspace*{-0.4cm}
\end{figure}

\subsection{Case 3: $0<p<1$}
In this case, the adversary partially allows node $n$ to receive incoming files from the gossip network. In (\ref{eqn:eclipse_vSk}), plugging $p=1$ in denominator gives lower bound $\frac{n-k}{n-k+1}\big(y_{k+1}+\frac{1}{k\lambda} + \frac{pv_n}{n-1} \big) <y_k$ and plugging $p=0$ in denominator gives upper bound $y_k<y_{k+1}+\frac{1}{k\lambda}+\frac{pv_n}{n-1}$, and together with techniques employed in cases $1$ and $2$, we can bound $v_1$ as
\begin{align}\label{eqn:upperbd_lowerbd_0p1}
    O(\log n) + \frac{pv_n}{2} <v_1<O(\log n) + pv_n 
\end{align}
Clearly how age scales at the infected node dictates the age scaling for the regular nodes in the rest of the network. For a fixed $p<1$, choosing $n\gg \frac{1}{1-p}$ can result in $v_n\approx \frac{\frac{1}{\lambda}+(1-p)v_1}{(1-p)}=O(1)+v_1$, which, when combined with (\ref{eqn:upperbd_lowerbd_0p1}) yields $O(\log n)$ age scaling for both $v_1$ and $v_n$. Hence, if the infected node is allowed to accept a small fraction of incoming packets from the network, then a large network can manage to curb the spread of stale files coming from the infected node by sustaining a low age at all nodes.\footnote{More generally, the adversary can increase timestamps of outgoing and incoming packets with probability $p$ and $q-1$, respectively, which changes (\ref{eqn:Xn_eclipse_adversary}) to $v_n=\frac{\frac{1}{\lambda}+(1-q)v_1}{\frac{1}{n}+(1-q)}$. Nevertheless, if $q=1$, similar to Case 1, $v_n=\frac{n}{\lambda}$, and since (\ref{eqn:upperbd_lowerbd_0p1}) remains unchanged, we get again $v_1=O(n)$. Likewise, when $q<1$, similar to Case 3, choosing $n\gg \frac{1}{1-q}$ again brings down age at all nodes to $O(n)$. Hence, the adversary can best worsen the age to $O(n)$ by not allowing incoming packets to infected node.}


\section{MITM Attack on Fully Connected Network}

In previous sections, the adversarial node was in direct contact with all other nodes due to fully connected nature of the network, and the adversary could raise the system age to $O(n)$ with $p=1$. Here, an interesting question to ask is if the network could do better if the adversary instead had access to only one node. To this end, we consider the network model of Fig.~\ref{fig:fully_conn_net_MITM_attack}, where the adversary, which we will refer to as node $A$, intercepts the updates to node $n$ coming from the source. In turn the adversary sends updates with rate $\lambda$, after changing the timestamps of every outgoing packet to current time, only to node $n$. 

Clearly the expected age at the adversary, denoted by $v_A$, scales as $O(n)$ since it is isolated from the gossip network and only receives updates from the source with rate $\frac{\lambda}{n}$. The two reset maps useful for our analysis are 
\begin{align} \label{eqn:mitm_X_NSkn_reset}
X_{N(S_k\cup\{n\})}' = \begin{cases} 
0, & i=0,j\in S_k\\
X_{N(S_{k+1}\cup\{n\})}, & i\in \mathcal{N}_R\backslash S_k,j\in S_k\cup\{n\}\\
X_A, & i=A,j=n\\
X_{N(S_k\cup\{n\})}, & \text{otherwise}
\end{cases}
\end{align}
and
\begin{align} \label{eqn:mitm_X_NSk_reset}
X_{N(S_k)}' = \begin{cases} 
0, & i=0,j\in S_k\\
X_{N(S_{k+1})}, & i\in \mathcal{N}_R\backslash S_k,j\in S_k\\
X_{N(S_k\cup\{n\})}, & i=n,j\in S_k\\
X_{N(S_k)}, & \text{otherwise}
\end{cases}
\end{align}
We claim $v_{S_k\cup\{n\}} \geq \frac{v_A}{2}$, a loose lowerbound that is trivially verified with induction as follows. Invoking (\ref{eqn:hespanha_eqn}) regarding (\ref{eqn:mitm_X_NSkn_reset}) for $k=n-1$ results in 
\begin{align}
    v_{S_{n-1}\cup\{n\}}=\frac{\frac{1}{\lambda}+  v_A}{\frac{n-1}{n} + 1} \geq O(1)+\frac{v_A}{2} \geq \frac{v_A}{2}
\end{align}
which verifies the claim for $k=n-1$. Next, we assume the claim holds for $k+1$, i.e., $v_{S_{k+1}\cup\{n\}}\geq \frac{v_A}{2}$, and verify for $k$. Invoking (\ref{eqn:hespanha_eqn}) regarding (\ref{eqn:mitm_X_NSkn_reset}) for $k\leq n-2$ and using $\frac{1}{\lambda}>0$ in the numerator and $\frac{k}{n}\leq 1$ in the denominator gives 
\begin{align} \label{eqn:nmitm_vskn_lb_vA_by_2}
    \!\!\!v_{S_k\cup\{n\}}=& \frac{\frac{1}{\lambda}+ \frac{(k+1)(n-1-k)}{n-1} v_{S_{k+1}\cup\{n\}} + v_A}{\frac{k}{n} +\frac{(k+1)(n-1-k)}{n-1} + 1}  \\
    \geq & \frac{\frac{(k+1)(n-1-k)}{n-1} v_{S_{k+1}\cup\{n\}}}{\frac{(k+1)(n-1-k)}{n-1} + 2} + \frac{v_A}{\frac{(k+1)(n-1-k)}{n-1} + 2}  \\
    \geq & \frac{\frac{(k+1)(n-1-k)}{n-1} \frac{v_A}{2}}{\frac{(k+1)(n-1-k)}{n-1} + 2} + \frac{\frac{2v_A}{2}}{\frac{(k+1)(n-1-k)}{n-1} + 2} \\
    =& \frac{v_A}{2}
\end{align}
Finally, we re-invoke (\ref{eqn:hespanha_eqn}) for (\ref{eqn:mitm_X_NSk_reset}) which results in 
\begin{align} \label{eqn:mitm_vSk_shs_stable_eqn}
    v_{S_k}=\frac{\frac{1}{k\lambda}+\frac{n-k-1}{n-1}v_{S_{k+1}}+\frac{v_{S_k\cup\{n\}}}{n-1}}{\frac{1}{n}+\frac{n-k-1}{n-1}+\frac{1}{n-1}}
\end{align}
Let $y_k=v_{S_k}\frac{n-k}{n-1}$, using $\frac{1}{n-1}\approx \frac{1}{n}$ for large $n$, (\ref{eqn:mitm_vSk_shs_stable_eqn}) becomes
\begin{align}
    y_k=&\frac{n-k}{n-k+1}\bigg( y_{k+1}+ \frac{1}{k\lambda} +  \frac{v_{S_k\cup{n}}}{n-1} \bigg) \\
    \geq&\frac{n-k}{n-k+1} y_{k+1}+ \frac{(n-k)v_{S_k\cup\{n\}}}{(n-k+1)(n-1)} 
\end{align}
Starting from $y_1=v_1$, we successively substitute for $y_2, y_3, \ldots, y_{n-1}$ and use $v_{S_k\cup\{n\}} \geq \frac{v_A}{2}$ to obtain
\begin{align} 
    v_1\geq& \frac{1}{n(n-1)}\sum_{k=1}^{n-1} (n-k)v_{S_k\cup{n}} \\
    \geq & \frac{v_A}{2n(n-1)}\sum_{k=1}^{n-1}(n-k) =\frac{v_A}{4} \label{eqn:mitm_v1_lb_va_by_4}
\end{align}
Hence, $v_1$ scales at least as $O(n)$ for all regular nodes. This result is far from intuitive, for it brings home the point how an adversary, with so little an effort as sending tampered packets to just one node, can bring down the freshness of an entire large gossip network.

\section{Numerical Results}

We simulate a fully connected network of size $n$ and allow it to gossip for a total time of $1000n$, choosing $\lambda =1$. 

\begin{figure}[t]
\centerline{\includegraphics[width=0.95\linewidth]{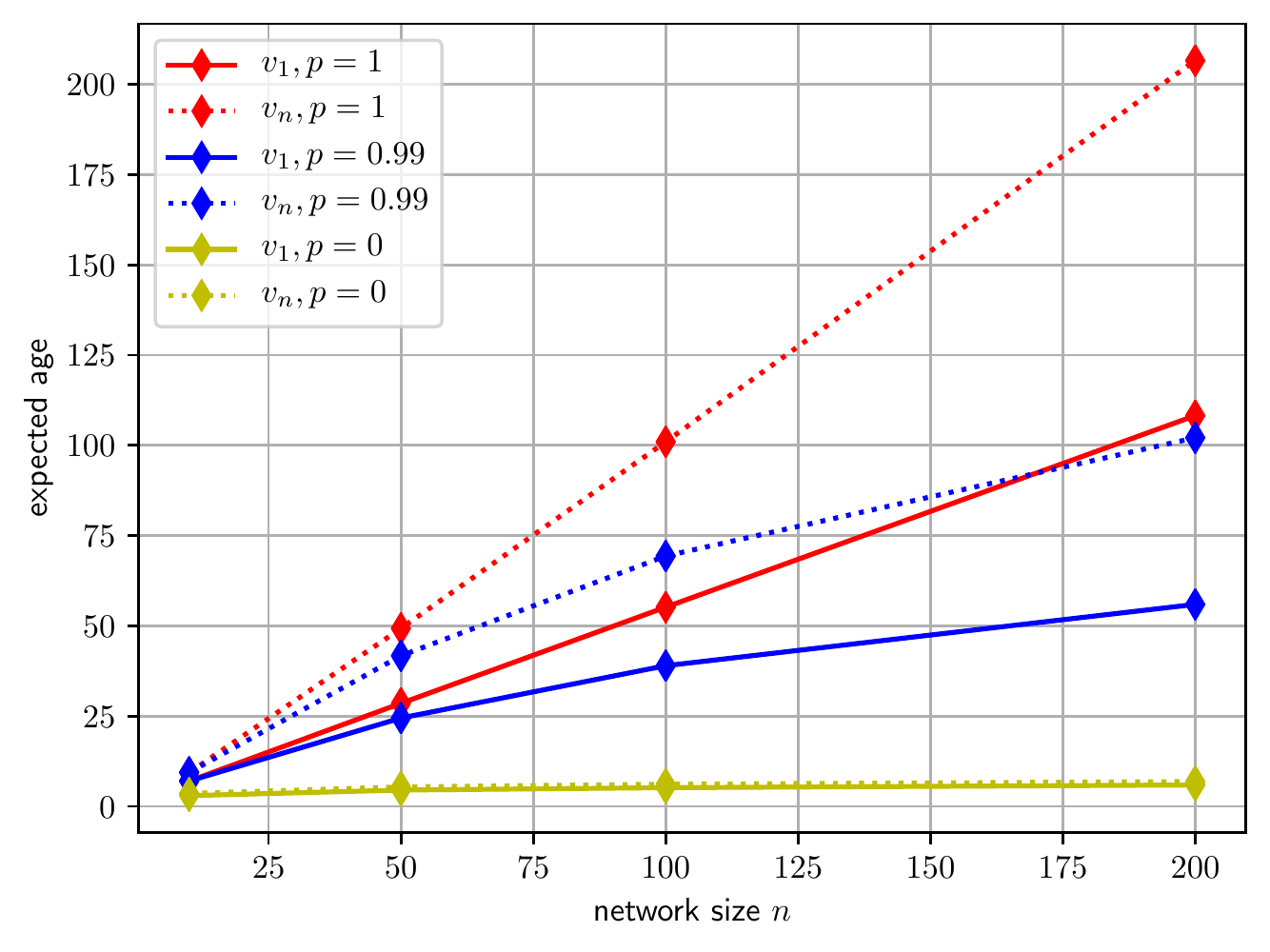}}
\caption{Node capture attack on fully connected network of $n$ nodes.}
\label{fig:graph_fully_conn_net_eclipse_attack}
\end{figure}

Fig.~\ref{fig:graph_fully_conn_net_eclipse_attack} shows the expected age at a regular node and the infected node for all three cases of $p$ of Section~\ref{sect:no-capture}. Focusing on the case $p=1$ shown in red color, the age at the infected node $v_n$ grows as $\frac{n}{\lambda}$ and the age at a regular node $v_1$ grows as $\frac{v_n}{2}=\frac{n}{2\lambda}$, as was analytically suggested in (\ref{eqn:eclipse_iteratively}) and (\ref{eqn:v1_vn_by_2_case1}). On the other extreme, $p=0$ gives logarithmic age scaling at all nodes, with the infected node age $v_n$ just slightly above the regular node age $v_1$, in accordance with (\ref{eqn:eclipse_vn_case2}). In the third case of $p=0.99$, which allows the infected node to accept $1 \%$ of incoming gossip, we observe that the infected node age $v_n$ initially begins to grow linearly but later starts to scale logarithmically for larger values of $n$ as $n$ becomes $n\gg \frac{1}{1-p}$. These results imply that the best course of action for the adversary should be to block all incoming traffic and actively send out outdated timestomped packets. 

\begin{figure}[t]
\centerline{\includegraphics[width=0.95\linewidth]{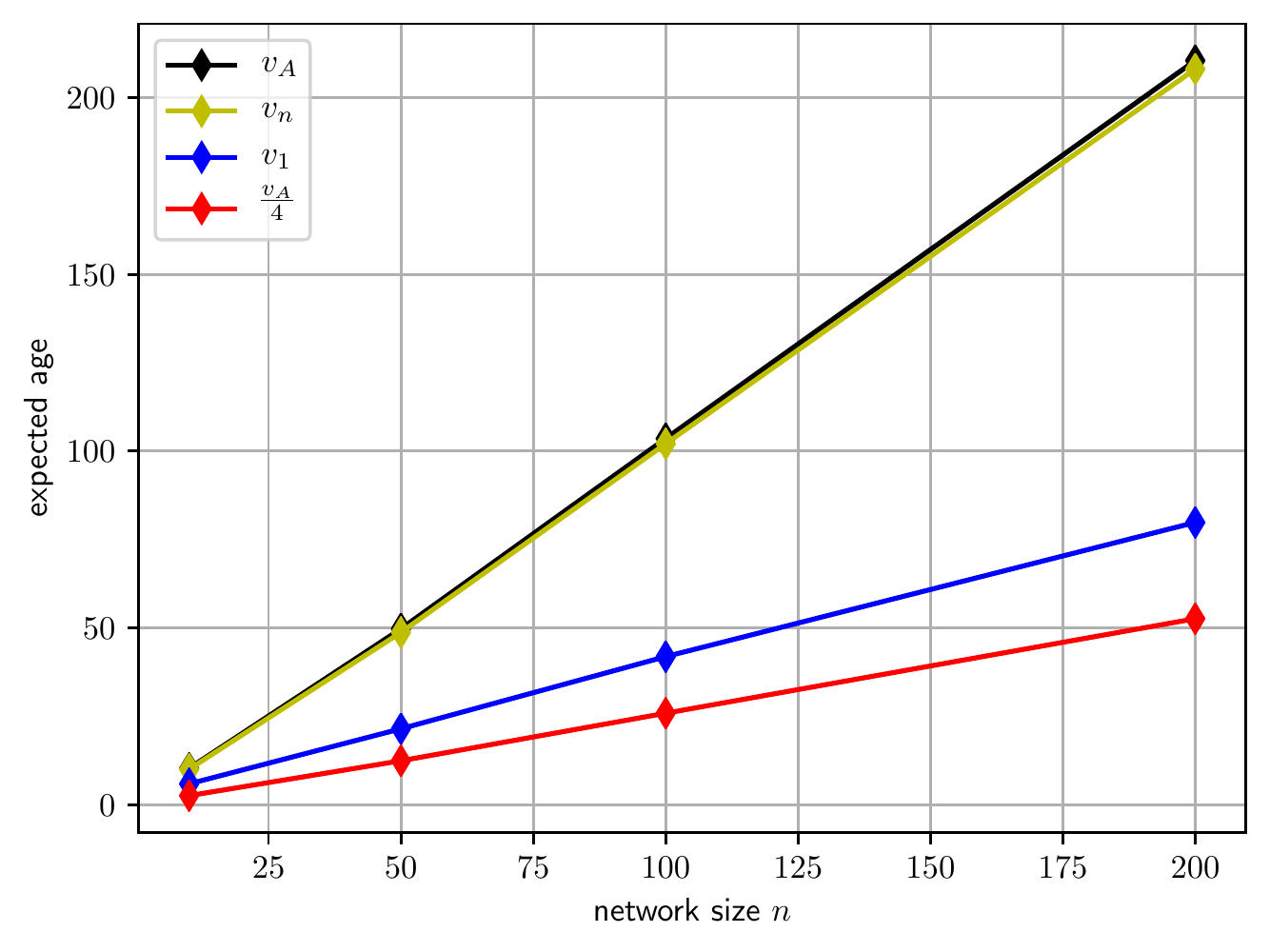}}
\caption{MITM attack on fully connected network of $n$ nodes.}
\label{fig:graph_fully_conn_net_MITM_attack}
\vspace*{-0.4cm}
\end{figure}

Fig.~\ref{fig:graph_fully_conn_net_MITM_attack} shows the expected age at different types of nodes when the adversary is positioned between the source and a node. The red line shows the lower bound $\frac{v_A}{4}$ of (\ref{eqn:mitm_v1_lb_va_by_4}), where the age at a regular node $v_1$ lies above this lower bound. Adversary age $v_A$  grows as $O(n)$ by virtue of being an isolated node. Finally, though (\ref{eqn:nmitm_vskn_lb_vA_by_2}) yields a loose lower bound of $\frac{v_A}{2}$, the graph shows that the age at the node that is in contact with adversary, $v_n$, closely follows adversary age $v_A$.

\section{Conclusion}
We studied the effects of timestomping attacks on the age of gossip in a large fully connected network. We showed that one infected node in such a network can increase the age at all other nodes from $O(\log n)$ to $O(n)$ through timestamp manipulation. Further, we showed that the optimal behavior for the adversary is to reset the timestamps of all outgoing packets to current time thereby disguising them as current packets and of all incoming packets to an outdated time to prevent their acceptance at the infected node. Additionally, we showed that if the adversary allows the infected node to accept even a very small fraction of the incoming packets from the network, then a large network can manage to curb the spread of stale files coming from the infected node and pull the network age back to $O(\log n)$. Lastly, we showed that if an infected node contacts only a single node instead of all nodes of the network, the system age can still be degraded to $O(n)$. 

\bibliographystyle{unsrt}
\bibliography{ref_priyanka}

\end{document}